\documentclass[floatfix,aps,prb,twocolumn,showpacs,superscriptaddress]{revtex4}
\usepackage{graphicx}
\usepackage[]{epsfig}
\usepackage{times}
\begin{document}

\title{Breakdown of `phase rigidity' and variations of the Fano effect in
 closed Aharonov-Bohm interferometers}
\author{Amnon Aharony}
\affiliation{Institute for Solid State Physics, University of Tokyo, 5-1-5
Kashiwanoha, Chiba 277-8581, Japan}
\affiliation{School of
Physics and Astronomy, Raymond and Beverly Sackler Faculty of
Exact Sciences,  Tel Aviv University, Tel Aviv 69978, Israel}
\affiliation{Physics Department, Ben Gurion University, Beer
Sheva 84105, Israel}
\author{Ora Entin-Wohlman}
\affiliation{Institute for Solid State Physics, University of Tokyo, 5-1-5
Kashiwanoha, Chiba 277-8581, Japan}
\affiliation{School of
Physics and Astronomy, Raymond and Beverly Sackler Faculty of
Exact Sciences,  Tel Aviv University, Tel Aviv 69978, Israel}
\affiliation{Physics Department, Ben Gurion University, Beer
Sheva 84105, Israel}
\author{Tomohiro Otsuka}
\affiliation{Institute for Solid State Physics, University of Tokyo, 5-1-5
Kashiwanoha, Chiba 277-8581, Japan}
\author{Shingo Katsumoto}
\email[]{kats@issp.u-tokyo.ac.jp}
\affiliation{Institute for Solid State Physics, University of Tokyo, 5-1-5
Kashiwanoha, Chiba 277-8581, Japan}
\author{Hisashi Aikawa}
\altaffiliation[Present address:]{\ Toshiba Corporation, Research
and Development Center, 1, Komukai Toshiba-cho, Saiwai-ku,
Kawasaki-shi, 212-8582 Kanagawa, Japan} \affiliation{Institute for
Solid State Physics, University of Tokyo, 5-1-5 Kashiwanoha, Chiba
277-8581, Japan}
\author{Kensuke Kobayashi}
\altaffiliation[Present address:]{\ Institute of Chemical
Research, Kyoto University, Gokasho Uji, Kyoto 611-0011, Japan}
\affiliation{Institute for Solid State Physics, University of
Tokyo, 5-1-5 Kashiwanoha, Chiba 277-8581, Japan}
\date{\today}

\begin{abstract}
Although the conductance of a closed Aharonov-Bohm interferometer,
with a quantum dot on one branch, obeys the Onsager symmetry under
magnetic field reversal, it needs not be a periodic function of
this field: the conductance maxima move with both the field and
the gate voltage on the dot, in an apparent breakdown of `phase
rigidity'.  These experimental findings are explained
theoretically as resulting from multiple electronic paths around
the interferometer ring. Data containing several Coulomb blockade
peaks, whose shapes change with the magnetic flux, are fitted to a
simple model, in which each resonant level on the dot couples to a
different path around the ring.
\end{abstract}

 \maketitle

\section{Introduction}

The mesoscopic Aharonov-Bohm interferometer (ABI) has been used
widely in attempts to measure both the magnitude and the phase of
the quantum transmission amplitude for an electron traversing a
quantum dot (QD). Many of these experiments have been done on the
{\it closed} ABI, where the QD is placed on one of the two paths
which surround an area which is penetrated by a magnetic flux
$\Phi$, and the two paths are connected to two reservoirs via only
two terminals.~\cite{MH,KK} In some experiments, the states on the
whole path replace the QD.~\cite{path} Unlike for the
multi-terminal {\it open} ABI, for small fluxes the conductance of
the closed interferometer turned out to be an {\it even and
periodic function of the Aharonov-Bohm (AB) phase}
$\phi=\Phi/\Phi_0$, with $\Phi_0=\hbar c/e$ and with the period
$\Delta\phi=2\pi$. Away from resonances of the transmission, and
for relatively small magnetic fields, the conductance of the
`ideal' closed ABI could be fitted to the simple (two-slit-like)
formula $G=A+B \cos\phi$, and therefore its maxima (and minima)
remained fixed at integer multiples of $\pi$, independent of the
gate voltage on the QD (which only affected the values of $A$ and
$B$). This phenomenon, called ``phase rigidity", has been accepted
as a landmark of the closed ABI.~\cite{yacoby2,hack} Closer to a
resonance, $G$ becomes a more complicated function of $\phi$,
which contains higher harmonics in the flux, but continues to
depend only on powers of $\cos\phi$ (and not of $\sin\phi$).
Indeed, the symmetry $G(\phi)=G(-\phi)$ is now well understood,
due to the Onsager relations.~\cite{onsager} The periodicity of
$G$ with $\phi$, and the resulting phase rigidity, have also been
reproduced theoretically, in models which describe both the paths
around the ABI ring and the leads to the reservoirs as being
one-dimensional (1D).~\cite{theory,W} However, although phase
rigidity obeys the Onsager symmetry, this rigidity does not really
{\it follow} from this symmetry. In fact, many of the measurements
at higher fluxes break phase rigidity.

The breakdown of phase rigidity in experiments on closed ABI's
shows up as deviations from the simple pure oscillation
$G=A+B\cos\phi$, even far away from resonances. Such deviations
already appeared in the pioneering work of Webb {\it et
al.},~\cite{webb} which demonstrated the AB oscillations in normal
metal rings. In these experiments (and in practically all the
other experiments mentioned above), the AB oscillations appear on
top of a background, whose slow variation with the magnetic flux
has been described as an {\it aperiodic fluctuation}, due to the
penetration of the magnetic field into conducting parts of the ABI
ring. Indeed, numerical simulations have shown that such
fluctuations do result from the fluxes which penetrate small areas
within the finite width of the ring, whose properties fluctuate
randomly.~\cite{stone, stoneYI} As noted by Imry,~\cite{book} such
fluctuations can be observed only at high fluxes, $\Phi \gg
\Phi_0/x$, where $x$ is the ratio of the area of the conducting
ring to the area of the hole inside it. As Stone and
Imry~\cite{stoneYI} note, this background may also contain beats.
However, we are not aware of a detailed theoretical analysis of
such beats, or of any other {\it periodic} aspect of the
deviations from a simple Aharonov-Bohm oscillation.

In the first part of this paper we concentrate on two aspects of
the breakdown of phase rigidity. First, the 'beats'. Figure 2 of
Ref. \onlinecite{webb} already showed two main peaks in the power
spectrum of the flux-dependent conductance: one at the
Aharonov-Bohm period and the other at a much smaller period
(higher field). Ignoring the aperiodic fluctuations, both periods
are clearly visible in the flux dependence of the conductance. The
ratio of the two periods is presumably related to the ratio
between the areas of the ring and of the hole, $x$. Similar
`beats' show up in practically all the experiments on closed
ABI's.~\cite{path} The second aspect concerns the location of the
conductance maxima and the related phase shift. Figure 4 of the
first reference in \onlinecite{KK} shows a contour plot of the
conductance versus the gate voltage and the magnetic field. In its
restricted sense, phase rigidity implies that (except close to
resonances) the maxima should be at fixed fields, namely on lines
parallel to the gate voltage axis. However, in the data (taken at
relatively high fields) these lines have a non-zero slope relative
to that axis. Similar slopes show up in many similar experimental
plots (see e.g. Fig. 4 in Ref. \onlinecite{keyser}, or Fig. 4 of
Ref. \onlinecite{sigrist}). A slow change of the location of the
maxima with increasing magnetic fields is visible even in the
original papers by Yacoby {\it et al.}~\cite{MH,yacoby2} Analysis
of the data within a finite narrow window of fields (away from
zero) would thus be described by $G \sim {\tilde A} +{\tilde
B}\cos(\phi+\delta)$, with a non-zero phase shift $\delta$, which
depends on both the flux and the gate voltage, apparently
contradicting phase rigidity. (This behavior holds only at large
fields; as the field goes to zero, $\delta$ also vanishes, in
accord with the Onsager requirement.) All of these papers also
exhibit a slow variation of the Aharonov-Bohm oscillation
amplitude ${\tilde B}$ with increasing field, which is related to
the `beats'. Below we present some additional experimental data,
and give a theoretical discussion of these observations.

The second part of this paper concerns the Fano shape of the
Coulomb blockade resonances. In the Coulomb blockade regime, the
QD exhibits a sequence of resonances as function of the gate
voltage, whenever another electron is added to its bound states.
The interference between these states and the continuum of the
electrons in the leads then results in the Fano
effect,~\cite{fano,jlt} which modifies the shape of these
resonances. These resonances are further modified once the QD is
placed in the ABI.~\cite{KK} In the experimental papers,~\cite{KK}
each resonance has been fitted to the `standard' Fano asymmetric
Breit-Wigner form, $G \propto |e+q|^2/(e^2+1)$, where $e$ is the
normalized distance of the gate voltage from its resonance value
and $q$ is the so-called `Fano asymmetry parameter', which can
become complex at non-zero magnetic fluxes, when time-reversal
symmetry is broken. Below we show that {\it all} the asymmetric
resonances follow from a {\it single} unified expression for the
transmission amplitude, and present fits to the data which
demonstrate the utility of this representation for many
resonances.~\cite{turk}

Section \ref{exp} presents some new data, taken from the same
mesoscopic closed ABI described in Ref. \onlinecite{KK}. Section
\ref{theory} then proceeds to describe a simple theory, which
takes account of the finite width of the ring. Finally, Sec.
\ref{fits} uses this theory to fit experimental data from the ABI.

\section{Experimental data}\label{exp}

\begin{figure}
\includegraphics[scale=1.2]{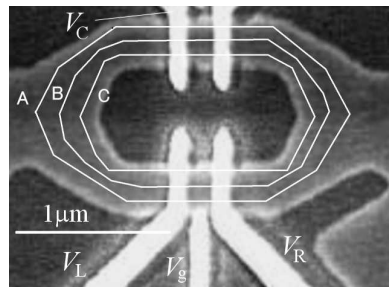}
\caption{Scanning electron micrograph of the ABI.
White regions are the Au/Ti gates and the corresponding
voltages $V_{\rm L}, V_{\rm g}, V_{\rm R}, V_{\rm C}$ are
indicated.
The contours A,
B and C indicate possible paths of the electrons.} \label{sample}
\end{figure}

For a quantitative discussion of the points listed above, we start
by presenting some new data, taken from the same sample described
in Refs. \onlinecite{KK}. As explained there, the closed
ring-shaped ABI, shown in Fig. \ref{sample}, was fabricated by wet
etching the 2DEG at an AlGaAs/GaAs heterostructure. Au/Ti metallic
gates define the QD and control the gate voltages on the QD
(sitting on the lower branch of the ring) and on the reference
(upper) branch. The sample was cooled by a dilution refrigerator,
the base temperature of which was 30mK though the electron
temperature measured from the line shape of the Coulomb
oscillation was around 100mK.

\begin{figure}
\includegraphics[width=1\linewidth]{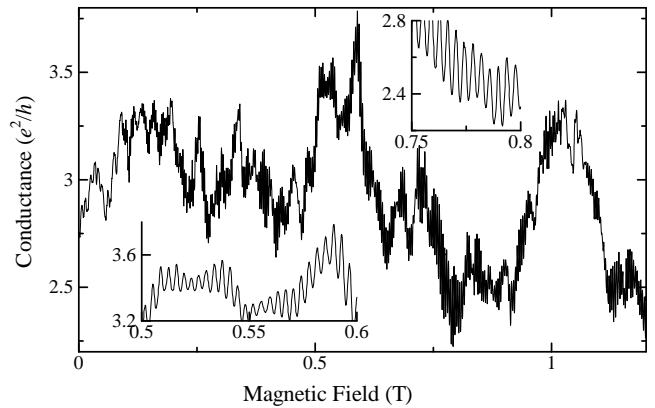}

\caption{The conductance of the ABI, with all the gates at zero
voltage. The insets are blowups in two different regions of
magnetic field.}
 \label{osci}
\end{figure}

\begin{figure}
(a) \includegraphics[width=0.75\linewidth]{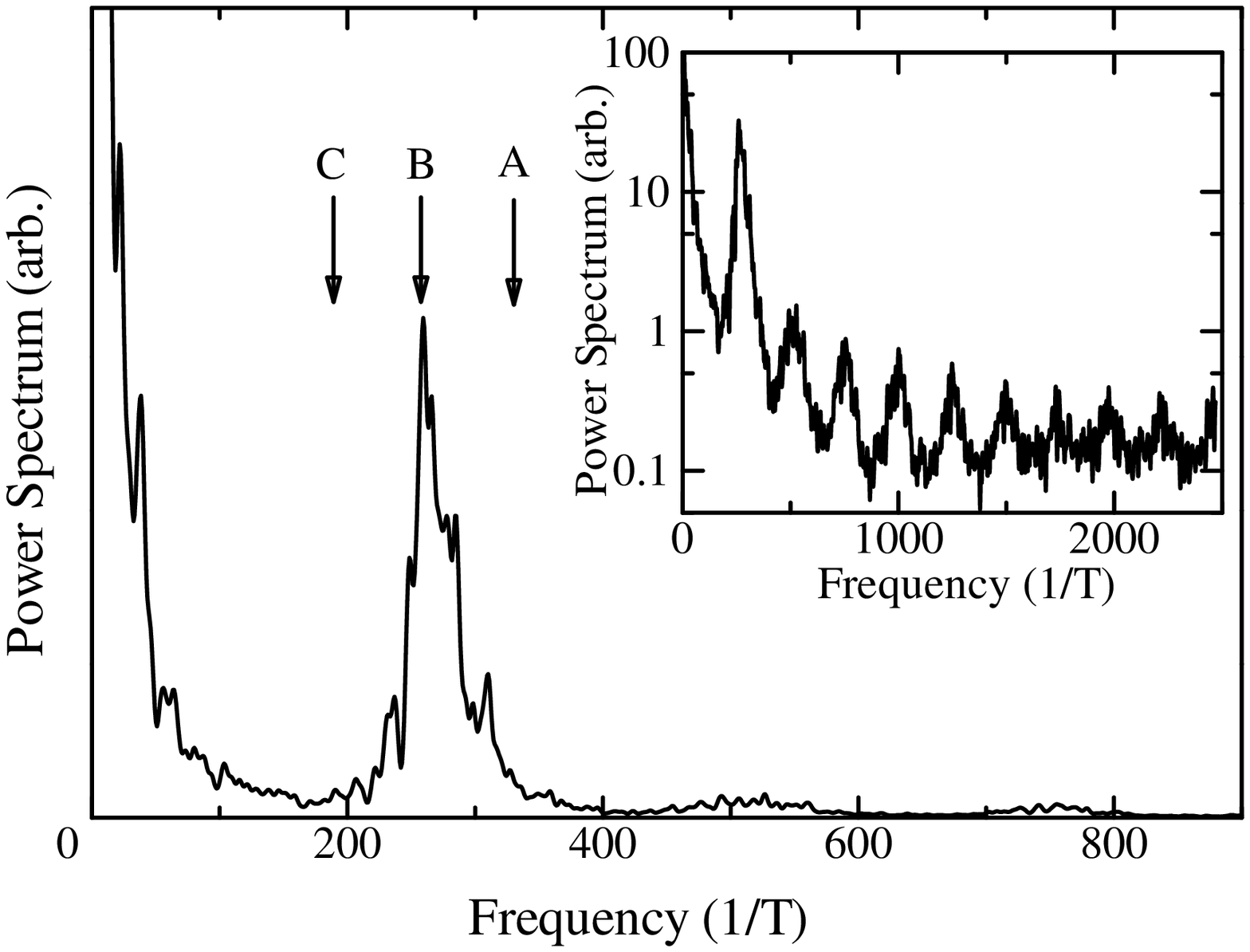}\\
(b) \includegraphics[width=0.75\linewidth]{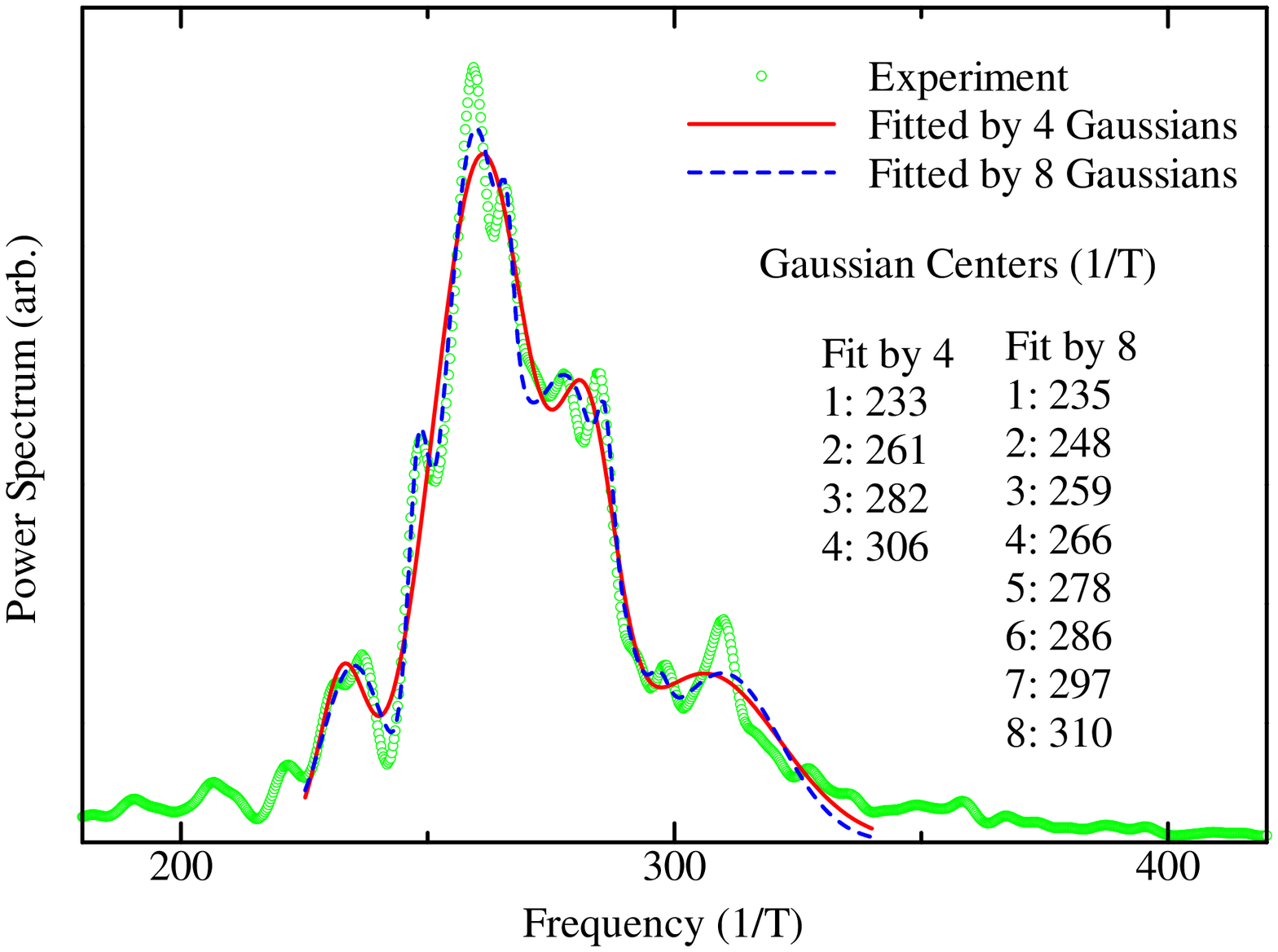}\\
(c) \includegraphics[width=0.75\linewidth]{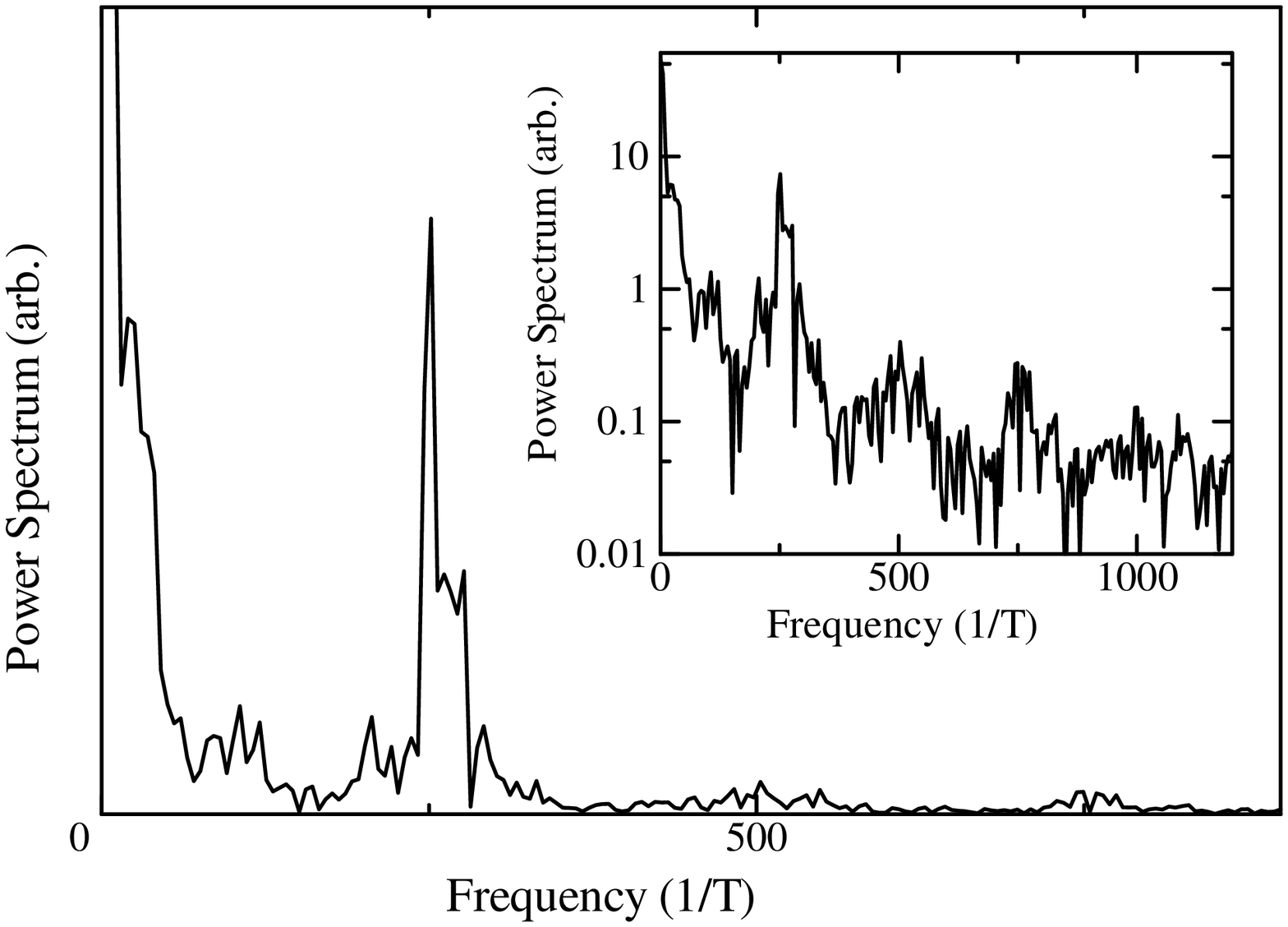}

\caption{Fourier power spectrum of the data in Fig. \ref{osci}
(see text). (a) FFT result for the entire field region in Fig.
\ref{osci}. A, B, C are the frequencies, which correspond to the
areas indicated by the contours A, B, C in Fig. \ref{sample}. The
inset is a log-plot of the same data, enhancing the higher
harmonics. (b) Results of the fitting to the main peak in (a) by 4
Gaussians (solind line) and by 8 Gaussians (broken line). (c)
Result of the same analysis as in (a) for the field region from
0.8T to 1T. } \label{FFT}
\end{figure}

All our data are found to be symmetric under $\phi \rightarrow
-\phi$, in agreement with the Onsager relations. In order to
investigate the nature of electronic paths in the ring geometry,
we first connected all the gates to ground. The sample was hence a
simple ring without a dot at this stage. Figure \ref{osci} shows a
typical flux dependence of the conductance through the ABI, for
these grounding conditions. The data clearly show the
Aharonov-Bohm oscillations, with a small period as demonstrated in
the insets. In addition,
apart from some aperiodic fluctuations, the data exhibit
oscillations on a larger scale. To quantify these oscillations,
Fig. \ref{FFT} shows the fast Fourier transform (FFT) of these
data. The top graph shows the FFT of all the data. Interestingly,
the results between $\sim$250/T and $\sim$280/T seem to contain
several separable peaks. The top graph also shows arrows for the
Aharonov-Bohm periods associated with the contours A, B and C in
Fig. \ref{sample}, indicating that all the frequencies in this
range can be associated with electron paths which surround the
ring between contours A and C. The inset in the top frame shows
the same FFT on a semi-logarithmic scale. It is interesting to
note that the data contain many higher harmonics, roughly at
integer multiples of the first one. The graph in the middle shows
fits to these high frequency data with four and with eight
Gaussians, confirming the impression that the fast oscillations
are dominated by only a few electron paths. The lower graph in
Fig. \ref{FFT} shows the FFT of the data with fields between 0.8T
and 1T. Interestingly, these data exhibit even fewer peaks,
implying that these restricted data can be described by `beats' of
a few neighboring frequencies.

\begin{figure}
\includegraphics[width=0.9\linewidth]{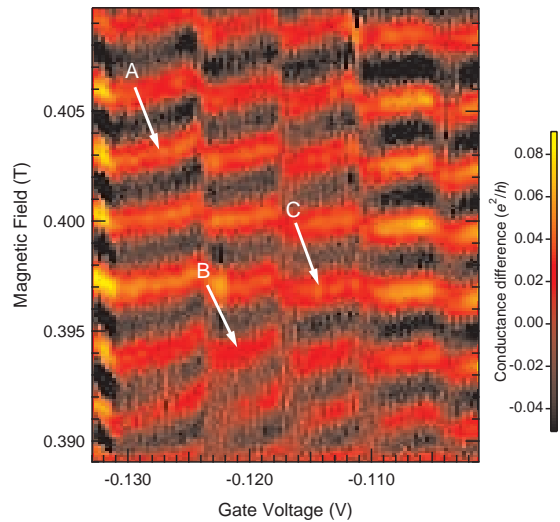}

\caption{False color image plots of the conductance, versus gate voltage $V_{\rm g}$ and
magnetic field. See the text for other voltages.
The linear baseline from 2.6 to 2.8 in $(e^2/h)$ was subtracted.
} \label{cplots}
\end{figure}

We have then formed a quantum dot in the lower branch, by applying
the gate voltages $-$0.255 V, $-$0.215 V for $V_{\rm L}$, $V_{\rm
R}$ in Fig. \ref{sample} respectively. We also reduced the
conductance of the reference arm, with $V_{\rm C}=-$0.26 V. Figure
\ref{cplots} is an image plot of the conductance as function of
the gate voltage $V_{\rm g}$ and the magnetic field. The slowly
varying background conductance, which is almost linear in this
region from 2.6 to 2.8 in conductance quantum units ($e^2/h$), was
subtracted. Figure \ref{cplots} clearly demonstrates the breakdown
of phase rigidity. Although a similar plot was presented in Ref.
\onlinecite{KK}, the present figure emphasizes several features
which we wish to discuss.

In Fig. \ref{cplots}, each resonance is characterized by a narrow
region (parallel to the field axis) through which the maxima (red
or bright) and minima (black or dark) interchange places. In the
ideal 1D model, between resonances the maxima remain fixed at
integer multiples of the flux unit, as follows from phase
rigidity. In addition, maxima and minima can suddenly interchange
between resonances, in a `phase lapse' which is attributed to a
Fano vanishing of the conductance. In contrast, the maxima in Fig.
\ref{cplots} never stay on lines parallel to the gate voltage
axis. Instead, they move continuously towards larger fields,
indicating a non-zero phase shift $\delta$. The variation of the
maxima can be characterized by three typical forms, indicated by
white arrows: in form A, the maximum moves almost linearly with
the gate voltage, so that $\phi$ changes by $\pi$, returning to
the value it had before the previous resonance. In form B, one
observes a fast change in the location of the maximum, about half
way between the resonances. Although reminiscent of the Fano jump
by $\pi$, this change has a finite width and seems continuous.
Finally, in form C the maximum moves by $\pi$ over the resonance,
but very soon it moves quickly back to its location before the
resonance, so that the line of this maximum remains almost
parallel to the gate voltage axis for a range of gate voltages.

More experimental data are presented below, in Sec. \ref{fits}.

\section{Theory}\label{theory}

As stated, most of the existing theoretical descriptions of the
ABI use models in which all the links are 1D. This means that the
ABI ``ring" surrounds a well defined unique area, which is
penetrated by a {\it single valued} magnetic flux $\Phi$. It is
this uniqueness that then results in the periodicity of $G$ with
$\phi$, resulting with phase rigidity. To explain the deviation of
our (and practically everyone else's) data from this periodicity,
we note that the real rings are {\it never} 1D (see Fig.
\ref{sample}).~\cite{webb} Due to the finite width of the ABI
ring, electrons which move on different paths within the ring
surround different areas, and therefore different magnetic
fluxes.~\cite{webb,stone,stoneYI,book} We demonstrate the
implications of these different fluxes by a simple tight-binding
model, which generalizes models used earlier to describe the ABI
in the Coulomb blockade regime.~\cite{theory,W} Our model is a
simplified version of that discussed in Refs. \onlinecite{stone}
and \onlinecite{stoneYI}: instead of the aperiodic random
fluctuations, we emphasize the {\it periodic} influence of the
flux through the conducting ring.

One basic difference between our device and that of Webb {\it et
al.}\cite{webb} is that our ring is semiconducting, while theirs
was metallic. In a semiconductor, we expect the electron to have
much fewer possible paths for traversing the ring. In view of
this, our model assumes that each resonance on the quantum dot
couples strongly only to a single wave function on the ring, and
that each such wave function can be associated with a distinct
area and therefore with a distinct magnetic flux.

\begin{figure}
\includegraphics[scale=0.7]{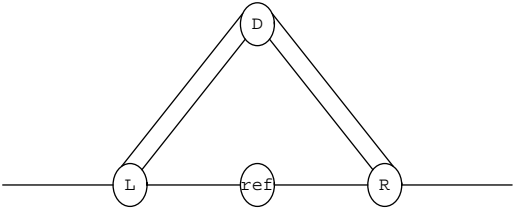}

\caption{The model for the ABI: all the tight-binding bonds on the
external 1D leads have the same real hopping energy $-J$. L and R
connect to each other indirectly, via $(N+1)$ branches (here we
show $N=2$): one connects to the reference site, via real hopping
coefficients $j_\ell$ and $j_r$, while each of the other $N$
connect from L and from R to the quantum dot (D), via hopping
coefficients $J_\ell(n)=J_\ell^0(n)e^{i\phi(n)}$ and a real
$J_r(n)$.} \label{ziur}
\end{figure}

 In our model,
shown in Fig. \ref{ziur}, the QD has $N$ equidistant resonances,
at effective energies $E_D(n)=\epsilon_d+(n-1)U,~~n=1,2,\dots,N$,
where $\epsilon_d$ is controlled by the gate voltage. (Here, $U$
represents the Coulomb repulsion, within a Hartree approximation,
and $n$ represents the $n$'th Coulomb blockade
resonance.~\cite{W}) Each resonant level is connected to the left
and right terminals (L and R) via separate single paths, which may
surround different areas.  Using gauge invariance, we include the
possibly different magnetic fluxes associated with each resonant
level in the respective hopping matrix elements,
$J_\ell(n)=J_\ell^0(n)e^{i\phi(n)}$ and real $J_\ell^0(n)$ and
$J_r(n)$.~\cite{comm} The phase $\phi(n)$ is associated with the
area surrounded by the wave function which couples to the $n$'th
resonance. For simplicity, the ``reference" site on the lower path
of the ABI ring has only one state, with energy $E_{ref}$, which
couples to the sites L and R via real hopping matrix elements
$j_\ell$ and $j_r$. As we show below, the interference between
these different paths gives a conductance which is an even
function of the magnetic field (all the $\phi(n)$'s are of course
proportional to this field), but has a more complex
field-dependence at high fields -- similar to the experimental
data.

 We next solve the model shown
in Fig. \ref{ziur}. The bonds on the semi-infinite external 1D
leads have hopping matrix elements $-J$ (below we measure all
energies in units of $J$).
 For an electron with
wave number $k$ and energy $\epsilon=-2J\cos ka$ ($a$ is the
lattice constant on the leads), we write the wave functions on the
left and right leads as $u(e^{ikna}+re^{-ikna})$ and $ute^{ikna}$
(below we present results in the center of the band, $ka=\pi/2$),
and then solve the $N+5$ linear equations in the $N$ wave
amplitudes at D, those at L, R and ``ref" and in $t$ and $r$. The
resulting transmission amplitude is~\cite{bih}
\begin{equation}
t=\frac{S_{\ell r} 2 i \sin ka} {(S_{\ell
\ell}+e^{-ika})(S_{rr}+e^{-ika})-|S_{\ell r}|^2}, \label{tt}
\end{equation}
where
\begin{eqnarray}
S_{xy}=\sum_{n}\frac{J_{x}(n)J_y(n)^\ast}{J[\epsilon-E_D(n)]}+\frac{j_x
j_y}{J(\epsilon-E_{ref})}
\end{eqnarray}
($x,y$ stand for $\ell,r$). Using the Landauer
formula,~\cite{land} we thus obtain the zero-temperature
conductance,
\begin{eqnarray}
\frac{G}{G_0}=T =\frac{4\sin^2ka|S_{\ell r}|^2}{\Bigl ||S_{\ell
r}|^2-(S_{\ell \ell}+e^{-ika})(S_{rr}+e^{-ika}) \Bigr
|^2},\label{GG}
\end{eqnarray}
where $G_0=2e^2/h$ is the basic conductance unit and $T\equiv
|t|^2$ is the transmission.

It is interesting to note that the transmission $T$ depends on the
magnetic field only via the combination
\begin{eqnarray}
|S_{\ell
r}|^2&=&\frac{j_\ell^2j_r^2}{J^2(\epsilon-E_{ref})^2}\nonumber\\
&+&\sum_{nn'}\frac{J_\ell^0(n)J_\ell^0(n')J_r(n)J_r(n')}
{J^2[\epsilon-E_D(n)][\epsilon-E_D(n')]}\cos[\phi(n)-\phi(n')]\nonumber\\
&+&\frac{2j_\ell j_r}{J(\epsilon-E_{ref})}
\sum_{n}\frac{J_{\ell}^0(n)J_r(n)}{J[\epsilon-E_D(n)]}\cos\phi(n).\label{SLR}
\end{eqnarray}
Indeed, away from a resonance one has $|S_{xy}|\ll 1$, the field
dependence of $G$ in Eq. (\ref{GG}) is dominated by the numerator
there, and the maxima (or minima) coincide with those of $|S_{\ell
r}|^2$. For $N=2$, this has the form
\begin{eqnarray}
|S_{\ell
r}|^2&=&A+B_1\cos\phi(1)+B_2\cos\phi(2)\nonumber\\
&+&C\cos[\phi(1)-\phi(2)].\label{Slr}
\end{eqnarray}
Since both $\phi(1)$ and $\phi(2)$ correspond to paths through the
`upper' branch of the ABI, it is reasonable to expect that these
two fluxes are quite close to each other. Assuming a ratio $(1+x)$
between the areas surrounded by the two paths, we denote
$\phi(1)=\phi$ and $\phi(2)=(1+x)\phi$, and then Eq. (\ref{Slr})
has the form
\begin{eqnarray}
|S_{\ell
r}|^2&=&A+[B_1+B_2\cos(x\phi)]\cos\phi\nonumber\\
&-&B_2\sin(x\phi)\sin\phi+C\cos(x\phi)\nonumber\\
&\equiv& {\tilde A}+{\tilde B}\cos(\phi+\delta),\label{slr2}
\end{eqnarray}
with
\begin{eqnarray}
\tan\delta&=&B_2\sin(x\phi)/[B_1+B_2\cos(x\phi)],\nonumber\\
{\tilde
B}&=&\sqrt{B_1^2+B_2^2+2B_1B_2\cos(x\phi)},\nonumber\\
{\tilde A}&=&A+C\cos(x\phi). \label{cof}
\end{eqnarray}
For $|x|\ll 1$, the parameters ${\tilde A},~{\tilde B}$ and
$\delta$ vary slowly with $\phi$, and therefore within a limited
window of magnetic fields the data look like in the two-slit open
ABI, with a phase shift $\delta$ which varies with $\phi$ and with
the gate voltage, represented by $\epsilon_d$. On larger field
ranges, Eq. (\ref{slr2}) exhibits beats, similar to those observed
experimentally. Note that the parameters in Eq. (\ref{cof}) may
change quite significantly as $\phi$ changes from zero to $\pi/x$.
Needless to say, the expression in Eq. (\ref{SLR}), and therefore
also $G$, is symmetric under field reversal [$\phi(n) \rightarrow
-\phi(n)$ for all $n$], as expected from the Onsager relations.

When $x=0$, one has $\delta=0$ and $\tilde A,~\tilde B$ remain
constant, as for the simple 1D model. In that limit, the maxima
between resonances remain fixed, with possible jumps by $\pi$ when
$(B_1+B_2)$ changes sign as function of the gate voltage
$\epsilon_d$.

\begin{figure}
\includegraphics[scale=0.6]{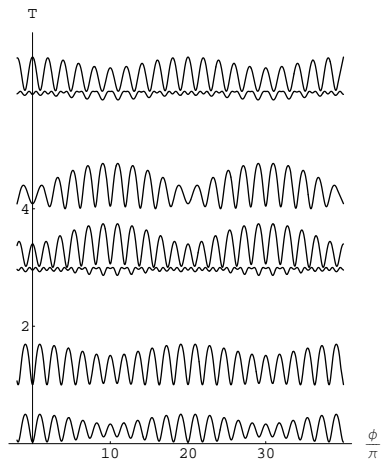}

\caption{Transmission through the closed ABI of Fig. \ref{ziur},
versus  $\phi$, at $\epsilon_d=-5.1,~-4,1,\cdots,0.9$ for the
parameters given in the text. Graphs are shifted up by 1 as
$\epsilon_d$ increases by 1.} \label{thosci}
\end{figure}

\begin{figure}
\includegraphics[scale=0.5]{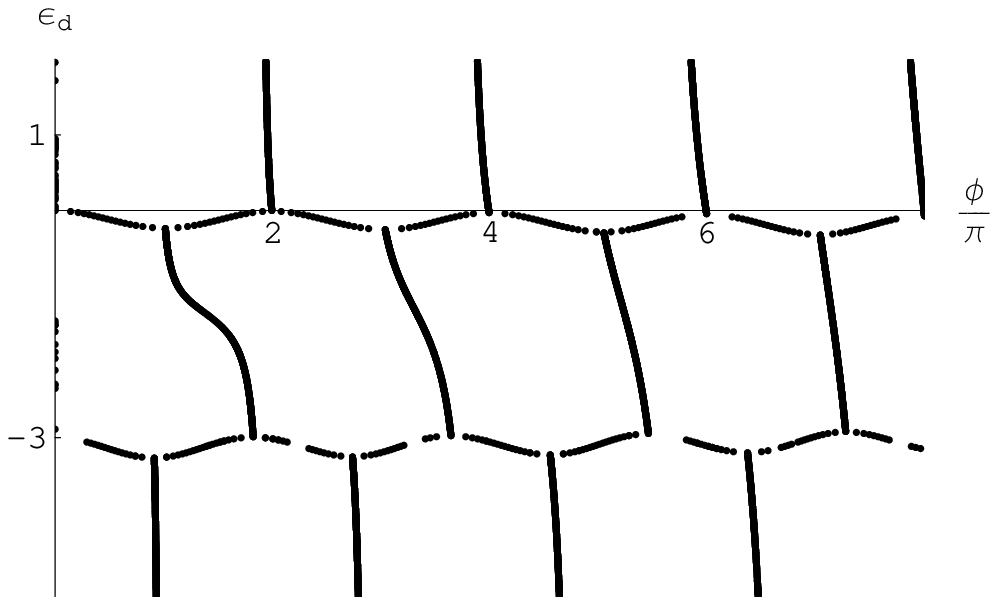}

\includegraphics[scale=0.5]{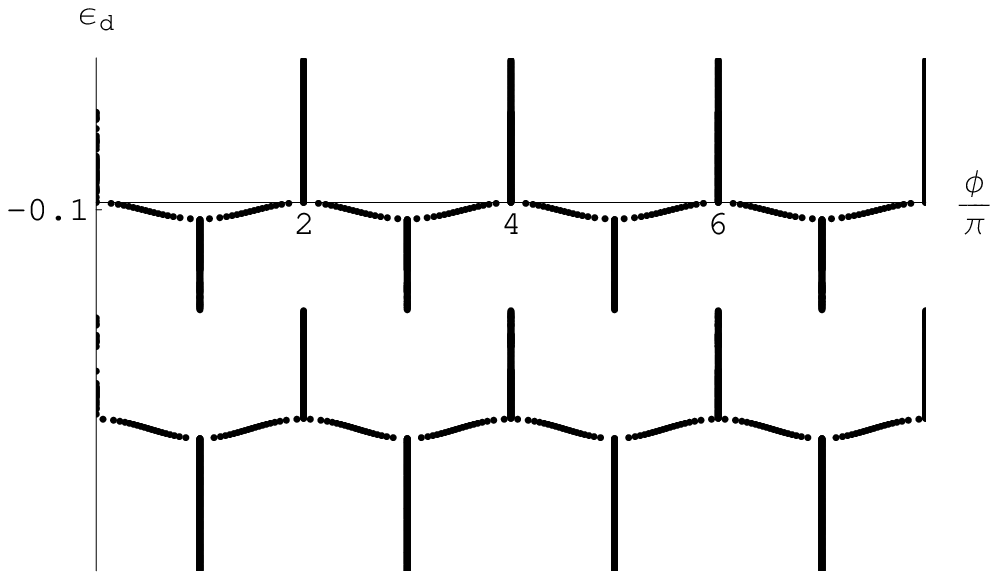}

\caption{The location of the maxima of $T$ in the
$\phi/\pi-\epsilon_d$ plane. Top: $x=0.1$. Bottom: $x=0$.}
\label{thcon}
\end{figure}

The above simple results change close to a resonance. To
demonstrate the full behavior of the conductance, we present an
example with $N=2$, with the parameters
$ka=\pi/2,~j_\ell=j_r=J_\ell^0(n)=J_r(n)=0.5J,~E_{ref}=J,~U=3J$
and $x=0.1$.  Figure \ref{thosci} shows the calculated
transmission, Eq. (\ref{GG}), for several values of the gate
voltage $\epsilon_d$. In addition to seeing the beats away from
resonances, we note the asymmetric beats closer to resonances. We
also note the gradual shifts in the maxima. These shifts are
highlighted in Fig. \ref{thcon} (top), which shows only the
locations of the maxima. This figure contrasts the behavior of the
maxima between the ideal 1D case, with $x=0$ (bottom) and the case
described above, $x=0.1$ (top). Note particularly the qualitative
change with increasing flux between the two resonances: for $x=0$
one observes a sharp `phase lapse', where the maxima jump from
even to odd multiples of $\pi$, due to an exact vanishing of the
conductance which results from the Fano interference between the
two resonances. These `lapses' are no longer sharp when $x \ne 0$:
in the range $1<\phi/\pi<2$ there appears a relatively fast change
of the maximum from around $\phi=2\pi$ down to $\phi \approx \pi$,
similar to form B in Fig. \ref{cplots}. However, as $\phi$
increases this `lapse' becomes smoother, and near $\phi=7\pi$ one
no longer sees such a `lapse' at all, as in form A in Fig.
\ref{cplots}.  At fluxes of order $\pi/x$, the interplay between
the two phases $\phi(1)$ and $\phi(2)$ destroys the exact
vanishing of the transmission between resonances, and thus also
destroys the `phase lapses'.

\begin{figure}
\includegraphics[scale=0.6]{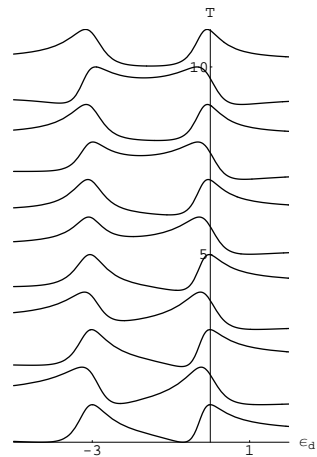}

\caption{Transmission through the ABI of Fig. \ref{ziur}, versus
gate voltage $\epsilon_d$, at $\phi=m\pi$. The graphs shift up by
1 as $\phi$ increases. } \label{thfano}
\end{figure}

The interplay between the two fluxes also affects the dependence
of the transmission on the gate voltage at fixed magnetic flux.
Figure \ref{thfano} shows this dependence for $\phi$ equal to
integer multiples of $\pi$. Although qualitatively similar, the
curves are not periodic in $\phi$, and one can see variations of
the Fano asymmetric shapes of the resonances with increasing flux.

Similar graphs also arise when one considers larger numbers of
resonances. As long as one has $\phi(n)=[1+x(n)]\phi$, with
$|x(n)|\ll 1$, the overall shapes of the graphs are found to be
similar to those presented above. Apparently, in the vicinity of a
specific resonance the results are mainly affected by neighboring
resonances, so that only a small number of different fluxes
participate in the `beats'.

\section{Fits to experimental data}\label{fits}

Equation (\ref{GG}) should represent a good approximation for any
sequence of Coulomb blockade peaks, with and without a magnetic
flux. Indeed, in Ref. \onlinecite{turk} this equation has been
applied to produce a reasonable imitation of the data found by
G\"ores {\it et al.}\cite{gores} for a mesoscopic single electron
transistor. The Fano asymmetry parameter for each resonance is in
fact determined by the influence of all the other  resonances, and
therefore there is no need for individual fits of $q$ for each
resonance and each set of parameters.

\begin{figure}[!b]
\includegraphics[width=0.8\linewidth]{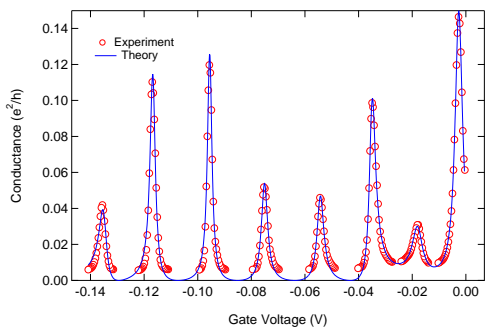}

\caption{Conductance from the pinched off ABI. Red (circles):
experiment. Blue (full line): fit to theory (see text). }
\label{NoRef}
\end{figure}

\begin{figure}
\includegraphics[width=0.8\linewidth]{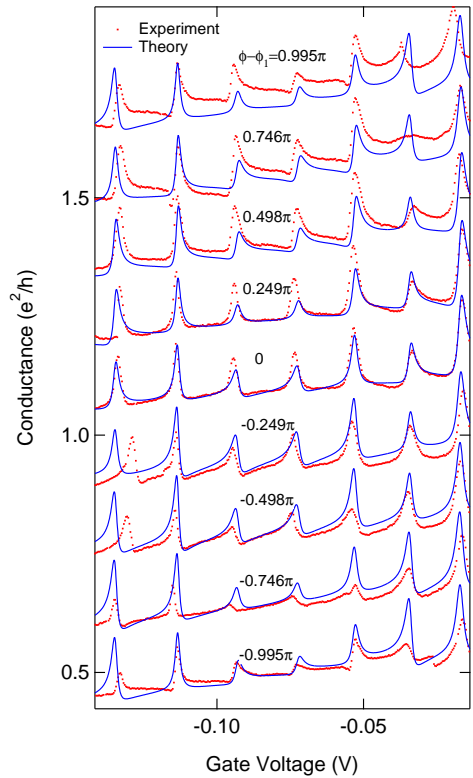}

\caption{Conductance from the ABI. Red (dots): experiment. Blue
(full lines): fit to theory (see text). Curves correspond to
fields $B$ between $0.9100$T and $0.9132$T, each graph is shifted
$0.15e^2/h$ } \label{compare}
\end{figure}

\begin{table*}
\begin{tabular}{crrrrrr}
\hline
\multicolumn{1}{c}{$n$}     &\multicolumn{1}{c}{$J_\ell^0(n)/\sqrt{eJ}$}               &\multicolumn{1}{c}{$J_r(n)/\sqrt{eJ}$}               &\multicolumn{1}{c}{$E_D(n)/e$}     &\multicolumn{1}{c}{$x(n)$}             &\\
                            &\multicolumn{1}{c}{$(\times10^{-2}\mathrm{V}^{1/2})$}&\multicolumn{1}{c}{$(\times10^{-2}\mathrm{V}^{1/2})$}&\multicolumn{1}{c}{$(\mathrm{V})$} &\multicolumn{1}{c}{$(\times10^{-2})$}  &\\
\hline
    1                       & $0.5886$                                            & $3.860$                                             & $0.0008$                          & 0                                     &\\
    2                       & $-3.761$                                            & $0.3806$                                            & $-0.0159$                         & $-5.913$                              &\\
    3                       & $3.708$                                             & $0.4211$                                            & $-0.0350$                         & $-0.069$                              &\\
    4                       & $3.876$                                             & $0.2988$                                            & $-0.0543$                         & $-0.085$                              &\\
    5                       & $3.716$                                             & $0.3073$                                            & $-0.0752$                         & $5.669$                               &\\
    6                       & $3.161$                                             & $0.4030$                                            & $-0.0956$                         & $-0.058$                              &\\
    7                       & $3.420$                                             & $0.4155$                                            & $-0.1168$                         & $-0.006$                              &\\
    8                       & $4.170$                                             & $0.2934$                                            & $-0.1353$                         & $0.069$                               &\\
\hline
\end{tabular}
\parbox{0.5\linewidth}{
\begin{tabular}{cc}
\hline
\multicolumn{1}{c}{$j_l/\sqrt{JE_{ref}}$}& \multicolumn{1}{c}{\ \ \ $j_r/\sqrt{JE_{ref}}$}\\
\hline
 $0.4224$ & \ \ \ $0.1003$\\
\hline
\end{tabular}

\vspace{4mm}
\begin{tabular}{cccc}
\hline
\multicolumn{1}{c}{$\phi_1$}&\multicolumn{1}{c}{$a$}&\multicolumn{1}{c}{$b$}&\multicolumn{1}{c}{$c$}\\
\multicolumn{1}{c}{$(\mathrm{rad})$}&\multicolumn{1}{c}{$(e^2/h)$}  &\multicolumn{1}{c}{$(e^2/h\mathrm{V})$}&\multicolumn{1}{c}{$(\mathrm{V})$} \\
\hline
 $1639.13$                          & $0.5392$                      & $0.7391$                              &   $-0.01768$                      \\
\hline
\end{tabular}
}

\caption[smallcaption]{The parameters of our fits to Eq.
(\ref{GG}). The conductance $G$ is also assumed to have a
background, $a+b V_g$, and the zero of $V_g$ for the ABI is
shifted by $c$ relative to that of the pinched off case. }
\label{tab}
\end{table*}

To demonstrate the effectiveness of Eq. (\ref{GG}), we have fitted
it to data from Ref. \onlinecite{KK} and to similar new data,
which exhibit a sequence of Coulomb blockade resonances of the
ABI, at various values of the magnetic flux. Here we chose a
sequence of measurements, done on the device shown in Fig.
\ref{sample} for gate voltages between $-0.142$ V and $-0.014$ V
and magnetic fields in the range $0.9100-0.9132$ T. Figures
\ref{NoRef} and \ref{compare} show fits to our data, using Eq.
(\ref{GG}). As explained in Ref. \onlinecite{KK}, our device
allows the pinch off of the reference path. Fitting Eq. (\ref{GG})
with $j_\ell=j_r=0$ to the pinched off data, we have determined
$E_D(n)$, $J_\ell^0(n)$ and $J_r(n)$, with $\phi(n)=0$. The
results of this fit are shown in Fig. \ref{NoRef}. We have then
added the reference path, and used the data from the ABI to
determine $j_\ell/\sqrt{E_{ref}}$, $j_r/\sqrt{E_{ref}}$ (we use
$\epsilon=0$, to represent electrons at the Fermi level, deep
inside the band) and the phases $\phi(n)\equiv [1+x(n)]\phi_1$ for
one particular value of the magnetic field, $B_1=0.9116$T. The
results of this fit appear as the fifth graph from the bottom in
Fig. \ref{compare}. Repeating this fit for $B_2=0.9100$T (
the curve at the bottom in Fig. \ref{compare}), using the same
parameters except for replacing $\phi_1$ by a fitted $\phi_2$, we
then found the coefficient $C$ in the relation $\phi=\phi_1+C
(B-B_1)$, $C=976.9\mathrm{T}^{-1}$.\cite{foot}
In addition, our
fit allows a background conductance which contains a component
linear in the gate voltage, in addition to a constant. The
resulting parameters are listed in Table \ref{tab}. Using these
parameters, we have then produced the theoretical curves for all
the other values of $B$, with no further adjustments (Fig.
\ref{compare}). The fits clearly capture all the qualitative
changes in the shapes of the resonances at different magnetic
fluxes. We find the results quite satisfactory, confirming the
assumptions of our theoretical model. We note that the fitted
parameters have practically equidistant resonances, with $U
\approx 0.02$ V. This Coulomb energy is consistent with the
capacitance and area of the quantum dot.

\begin{figure}
\includegraphics[width=0.9\linewidth]{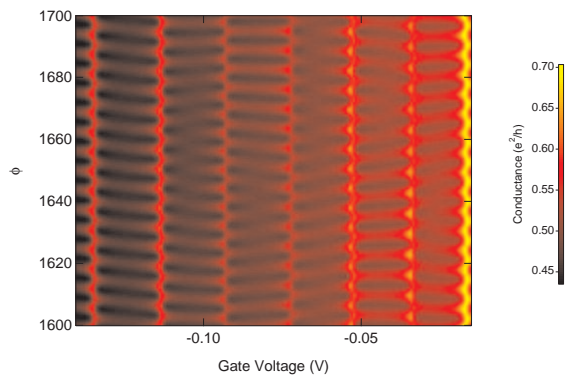}

\caption{Theoretical contour plot of the conductance. }
\label{image}
\end{figure}

Having obtained the parameters in Table \ref{tab}, we have then
plotted the theoretical contour plot of the conductance, see Fig.
\ref{image}. Qualitatively, this figure is similar to Fig.
\ref{cplots} (taken with different parameters): the maxima move
continuously with the magnetic flux and with the gate voltage,
imitating typical experimental data.

\section{Conclusion}

In this paper we have concentrated on the periodic effects of
having a closed ABI with a ring which has a finite width. For a
semiconductor ring, we argue that each resonance on the quantum
dot can be associated with a single magnetic flux, which
penetrates the wave function in the ring which couples to that
resonance. A simple theoretical formula then captures all the
qualitative features observed in many experiments.

We acknowledge very helpful discussions with Y. Imry, and support
from the Israel Science Foundation, under grant No. 1566/04, and
from a Grant-in-Aid for Scientific Research by the Japanese
Minstry of Education, Science, Sports and Culture.  AA and OEW are
also grateful to the ISSP for the warm hospitality.


\begin{references}
\bibitem{comm} We neglect here the magnetic width of the
wave functions, which may be important at high fields.
\bibitem{foot} For the fits at $B=B_1$, each $x(n)$ is determined only up to the
addition of $2\pi m(n)/\phi_1$, with integer $m(n)$.
We have also added small shifts to the parameters of peaks 1 and 2
to compensate some instability in the dot.
\end{references}
\end{document}